\documentclass{iopconfser}

\usepackage[style=numeric,backend=bibtex,sorting=none,url=false]{biblatex}
\bibliography{bibfile}
\DeclareFieldFormat{issn}{}
\renewbibmacro{in:}{}
\DeclareFieldFormat[article]{volume}{\textbf{#1}}
\DeclareFieldFormat
  [article]{title}{}	
\DeclareBibliographyDriver{article}{%
  \printnames{author}\newunit\newblock
	 \printfield{year}\setunit{\space}\newblock
  \printfield{title}\setunit{\addcomma\space}\newblock
  \printfield{journaltitle}\setunit{\addcomma\space}%
  \printfield{volume}\setunit{\addspace}%
  \printfield{number}\setunit{\addcomma\space}%
  \printfield{pages}\newunit\newblock
	\newunit
	\finentry}%

\DeclareBibliographyDriver{book}{%
\printnames{author}\setunit{\space}%
\printfield{year}\setunit{\addcomma\space}%
\printfield{title}\setunit{\addcomma\space}%
\printlist{publisher}%
\newunit
\finentry}

\usepackage{amsmath}
\renewcommand{\vec}[1]{\mathbf{#1}}
\usepackage{braket}

\begin{document}

\title{From Symmetry and Reduction to Physically Meaningful Relational Observables in Many-Body Quantum Theory}

\author{Ville J. Härkönen}

\affil{Computational Physics Laboratory, Tampere University, P.O. Box 692, FI-33014, Tampere, Finland  \\
Helsinki Institute of Physics, P.O. Box 64, FI-00014 Helsinki, Finland}
%\affil{$^2$Department, Institution, City, Country}

\email{ville.j.harkonen@gmail.com}

\begin{abstract}
We consider symmetries and reduction in non-relativistic many-body quantum mechanics, with the aim of identifying physically meaningful observables in systems such as molecules and crystalline solids. To this end, we propose a unified framework based on two additional postulates supplementing the standard quantum-mechanical formalism. For stable systems, the physically relevant states are normalizable stationary states, while physically meaningful observables are required to be invariant under a selected subgroup of the symmetry group and under Galilean boosts. In addition, we postulate the existence of a map from the set of all observables allowed by quantum mechanics to the corresponding invariant physically meaningful observables. The originality of the present work does not lie in specific reductions, but in the unified framework that connects symmetry reduction and relational many-body quantum theory. We interpret entities like superselection rules and quantum reference frames as important parts of the postulated process of obtaining the physically meaningful relational description. In particular, the requirement of Galilean-boost invariance added strengthens the criterion for physical observability by excluding quantities that depend on the choice of inertial frame. An important consequence of the postulates is that in the considered cases every physically meaningful observable necessarily depends on more than one non-invariant observable, the latter being typically associated with degrees of freedom assigned to a single particle. The postulates thus lead to theories that are well aligned with the literature on reduction and the description of molecules, while at the same time being consistent with the relational interpretation of quantum mechanics, according to which the complete physical description of a system is defined only relative to other systems.
\end{abstract}

\section{Introduction}
\label{Introduction}

Symmetries are among the most central aspects of physical theories, including quantum mechanics \cite{Dirac-PrinciplesOfQM-1958}. Symmetries \cite{Wigner-SymmetriesAndReflections-1967} reveal the conserved quantities \cite{Noether-InvarianteVariationsprobleme-1918}, define forces of a particular theory \cite{Yang-Mills-ConservationOfIsotopicSpinAndIsotopicGaugeInvariance-1954}, distinguish different phases of matter \cite{Landau-OnTheTheoryOfPhaseTransitions-1937}, simplify the computations \cite{Cornwell-GroupTheoryInPhysics-1997} and set constraints on the laws of nature \cite{Weinberg-TheQuantumTheoryOfFieldsVolumeIFoundations-1995}. Here we consider symmetries mostly from the point of view of them dictating which observables in quantum mechanical many-body theories are physically meaningful and the process of making non-meaningful physical observables to meaningful ones. This process is often called reduction, it has been known for a long time \cite{Newton-PhilosophiaNaturalisPrincipiaMathematica-1687,Lagrange-MecaniqueAnalytique-1788,Hamilton-OnaGeneralMethodInDynamics-1834,Jacobi-VorlesungenUberDynamik-1842,Poincare-LesMethodesNouvellesDeLaMecaniqueCeleste-1892}, especially in molecular physics \cite{Born-OppenheimerAdiabaticApprox-1927,Eckart-SomeStudiesConcerningRotatingAxesAndPolyatomicMolecules-PhysRev.47.552-1935,Nielsen-TheVibrationRotationEnergiesOfMolecules-RevModPhys.23.90-1951,Wilson-MolecularVibrationsTheTheoryOfInfraredAndRamanVibrationalSpectra-1955,Watson-SimplificationOfTheMolecularVibrationRotationHamiltonian-1968,Woolley-QuantumDynamicsOfMoleculesTheNewExperimentalChallengeToTheorists-1980,Sutcliffe-TheDecouplingOfElectronicAndNuclearMotions-2000,Kreibich-MulticompDFTForElectronsAndNuclei-PhysRevLett.86.2984-2001,Watson-TheMolecularVibrRotatKineticEnergyOperForGeneralInternalCoord-2004,Bouzas-ManyBodyHamiltoniansInImplicitlyDefinedFrames-2005} and nuclear physics \cite{Wheeler-MolecularViewpointsInNuclearStructure-PhysRev.52.1083-1937,Peierls-TheCollectiveModelOfNuclearMotion-1957,Lipkin-CollectiveMotionInManyParticleSystems-1960}. The modern reduction theory is highly developed field \cite{Rieffel-InducedRepresentationsOfCalgebras-1974,Landsman-MathematicalTopicsBetweenClassicalAndQuantumMechanics-1998,Littlejohn-GaugeFieldsInTheSeparOfRotatAndIntMotionsIntheNbodyProb-RevModPhys.69.213-1997,Marsden-MechanicalSystemsSymmetryAndReduction-2009,Marsden-IntroductionToMechanicsAndSymmetry-2013}, but still there are some approaches to quantum many-body theory where reduction is not addressed widely. Namely, most of the field theoretic many-body Green's function theories of Coulomb problem of electrons and nuclei are formulated unreduced \cite{Baym-field-1961,Giustino-ElectronPhononInteractFromFirstPrinc-RevModPhys.89.015003-2017,Stefanucci-InAndOutOfEquilibriumAbInitioTheoryOfElectronsAndPhonons-PhysRevX.13.031026-2023}, or in other words, absolute \cite{Harkonen-BeyondBornOppenheimerGreensFunctionTheoriesAbsoluteAndRelational-2025}. Consequently, we can show that these formulations can suffer from issues related to symmetries, the lack of reduction, and this has motivated for the development of reduced (or relational) formulation of the Coulomb problem of electrons and nuclei in terms of Green's functions \cite{vanLeeuwen-FirstPrincElectronPhonon-PhysRevB.69.115110-2004,Harkonen-ManyBodyGreensFunctionTheoryOfElectronsAndNucleiBeyondTheBornOppenheimerApproximation-PhysRevB.101.235153-2020,Harkonen-ExactFactorizationOfTheManyBodyGreensFunctionTheoryOfElectronsAndNuclei-PhysRevB.106.205137-2022}. The issues of the unreduced Green's function theories have been further discussed recently and we have connected \cite{Harkonen-BeyondBornOppenheimerGreensFunctionTheoriesAbsoluteAndRelational-2025} reduction to more recent developments like relational quantum mechanics \cite{Barbour-RelationalConceptsOfSpaceAndTime-1982,Rovelli-RelationalQuantumMechanics-1996} and quantum reference frames \cite{Aharonov-ChargeSuperselectionRule-PhysRev.155.1428-1967,Aharonov-QuantumFramesOfReference-PhysRevD.30.368-1984,Poulin-ToyModelForARelationalFormulationOfQuantumTheory-2006,Dickson-AviewFromNowhereQuantumReferenceFramesAndUncertainty-2004,Loveridge-SymmetryReferenceFramesAndRelationalQuantitiesInQuantumMechanics-2018}.

The purpose of this work is to summarize reduction and provide new perspectives on why reduction is needed in non-relativistic many-body quantum systems in general, and thus also in many-body Green's function theories. We also give a better justification why reduced formulation is needed given that the system has certain symmetries and pinpoint why reduced theories can be seen as relational. To make this concrete, we set two new observable-based postulates needed to make a generic many-body quantum theory of normalizable states to provide physically meaningful observables. This enables a definition of reduced and thus relational many-body quantum theories, including many-body Green’s functions theories, on the algebra of relational observables. These considerations connect the reduced formulations of the many-body problems to other modern developments related to quantum reference frames \cite{Aharonov-ChargeSuperselectionRule-PhysRev.155.1428-1967,Aharonov-QuantumFramesOfReference-PhysRevD.30.368-1984,Bartlett-ReferenceFramesSuperselectionRulesAndQuantumInformation-RevModPhys.79.555-2007,Giacomini-QuantumMechanicsAndTheCovarianceOfPhysicalLawsInQuantumReferenceFrames-2019} and relational quantum mechanics \cite{Zurek-EnvironmentInducedSuperselectionRules-PhysRevD.26.1862-1982,Rovelli-RelationalQuantumMechanics-1996,Poulin-ToyModelForARelationalFormulationOfQuantumTheory-2006,Loveridge-SymmetryReferenceFramesAndRelationalQuantitiesInQuantumMechanics-2018,Calosi-RelationalQuantumMechanicsAtTheCrossroads-2024}. Namely, we have claimed that the reduced theories can be seen as relational theories while the unreduced ones as absolute \cite{Harkonen-BeyondBornOppenheimerGreensFunctionTheoriesAbsoluteAndRelational-2025} and we try to make this connection stronger in this work.

This work is organized as follows. In Sec. \ref{Symmetry} we discuss symmetries in general non-relativistic problems and justify the need for reduction. We also introduce two postulates that can be used to transform an absolute many-body quantum theory to a relational one. Based on our postulates, we outline the reduction process to make a general theory to provide physically meaningful observables in Sec. \ref{Reduction}. After these general considerations, we discuss the many-body Green's function theories of Coulomb problem of electrons and nuclei in Sec. \ref{ManyBodyGreensFunctionTheories} and conclude our discussion in Sec. \ref{Conclusions}.

\section{Symmetry}
\label{Symmetry}

Our main interest here is the many-body problem of interacting electrons and nuclei. In particular, we are interested in stable bound states of these many-body systems. By stable we mean that the systems properties persist as a function of time. Therefore the objects of interest are the stationary states \cite{Dirac-PrinciplesOfQM-1958}, which are the eigenstates of the Hamiltonian, and which are normalizable (square integrable in some representation), a necessary condition for the probabilistic interpretation of quantum mechanics. We first consider rather general properties of the non-relativistic quantum mechanical $N$-body problem of a general form defined by the Hamiltonian operator
\begin{equation}
\hat{H} = \hat{T} + \hat{V},
\label{eq:SymmetryEq_1}
\end{equation}
where we denote all the position operators as $\hat{\vec{z}} \equiv \hat{\vec{z}}_{1}, \ldots, \hat{\vec{z}}_{N}$, the potential energy operator as $\hat{V} = \hat{V}\left(\hat{\vec{z}}\right)$ and the kinetic energy operator as $\hat{T} = \sum^{N}_{j = 1} {\hat{\vec{p}}^{2}_{j}}/{2 m_{j}}$. A special case of this is the Coulomb problem of electrons and nuclei. We omit spin degrees of freedom in this work. One important feature of the problem is symmetry, which are represented by unitary and anti-unitary transformations such that
\begin{equation}
\hat{U}^{\dagger}\left(g\right) \hat{H} \hat{U}\left(g\right) = \hat{H}, \quad g \in G,
\label{eq:SymmetryEq_2}
\end{equation}
where $G$ is the symmetry group. For our current discussion, particularly relevant are those $g \in G$ which are related to symmetries of physical space, that is, invariance of the Hamiltonian under translations and/or rotations of all particle coordinates. In particular, when we translate or rotate the whole many-body system comprising the whole universe, meaning that it has the aforementioned global symmetries, the meaningful physical properties must be invariant. Namely, given an observable $\hat{O}$, for it to be physically meaningful, it must be invariant under the symmetry transformations corresponding to translations and rotations. This reflects the fact that global translations and rotations of the entire system do not lead to physically distinguishable configurations, that is, the internal properties of a molecule or solid cannot depend on the choice of description, where the very same system is located as a whole in space. Another way to state this is to view the unitary symmetry transformations corresponding to translations and rotations as maps between frames of reference. Thus, the invariance condition amounts to requiring that physics is the same in all frames related by translations and/or rotations. The question then remains, what are the invariant internal observables and which are not. We can already mention at this point, as we will shortly see, that these are not in general the observables in terms of which $\hat{H}$ is written and has the symmetries of Eq. \eqref{eq:SymmetryEq_2}. These aspects have been noted in various disciplines and are stated with different terminology, as pointed out recently \cite{Harkonen-BeyondBornOppenheimerGreensFunctionTheoriesAbsoluteAndRelational-2025}. In quantum reference frames and superselection rules related literature, it is stated such that the translational symmetry leads to a superselection rule for momentum that dictates that absolute positions cannot be measured \cite{Aharonov-ChargeSuperselectionRule-PhysRev.155.1428-1967,Page-EvolutionWithoutEvolutionDynamicsDescribedByStationaryObservables-PhysRevD.27.2885-1983,Aharonov-QuantumParadoxesQuantumTheoryForThePerplexed-2005}. Similar reasoning applies when system has rotational symmetry, the absolute orientation is not observable \cite{Page-EvolutionWithoutEvolutionDynamicsDescribedByStationaryObservables-PhysRevD.27.2885-1983}. More generally, it has been claimed in relational quantum mechanics related literature that absolute quantities themselves are not meaningful, but the properties of a quantum system must always be characterized with respect to other systems \cite{Zurek-EnvironmentInducedSuperselectionRules-PhysRevD.26.1862-1982,Rovelli-RelationalQuantumMechanics-1996}.

These non-invariant observables also pass on the mentioned issues also to other observables written as a function of them. This can be seen by considering a translational invariant Hamiltonian as an example. For this Hamiltonian $\hat{U}^{\dagger}\left(\vec{a}\right) \hat{H} \hat{U}\left(\vec{a}\right) = \hat{H}$, where $\hat{U}\left(\vec{a}\right)$ is a symmetry unitary corresponding to a translation of all particles by an amount $\vec{a}$. The eigenstates of this Hamiltonian can be obtained from the time-independent Schrödinger equation
\begin{equation}
\hat{H} \ket{\Psi} = E \ket{\Psi}.
\label{eq:SymmetryEq_3}
\end{equation}
However, from the translational symmetry it can be shown that this Hamiltonian has purely continuous spectra and the states are not normalizable (not square integrable) \cite{Reed-MethodsOfModMathPhys-Vol4-1978,Sutcliffe-TheDecouplingOfElectronicAndNuclearMotions-2000}, thus not allowing a probability interpretation. Thus such eigenstates of a Hamiltonian as such cannot describe states of matter like bound states of molecules and crystals. Thus a translationally invariant Hamiltonian does not admit physically meaningful (Hilbert-space) eigenstates we are interested here.

There is one more case we want to discuss before setting up postulates to be used in obtaining physically meaningful observables. Namely, central to our discussion are the position and momentum variables in terms of which the Hamiltonian of Eq. \eqref{eq:SymmetryEq_1} is written. For definiteness, we denote the set of all observables by $\mathcal{A}$ and the set of all physically meaningful observables by $\mathcal{A}_{phys} \subset \mathcal{A}$. We assume for the moment that the Hamiltonian has translation and rotation symmetries. The Hamiltonian itself belongs to $\mathcal{A}$, but still does not necessarily admit physically meaningful eigenstates as these states are not normalizable and thus the Hamiltonian does not belong to $\mathcal{A}_{phys}$. The position operators transform as
\begin{equation}
\hat{U}^{\dagger}\left(\vec{a}\right) \hat{\vec{z}}_{j} \hat{U}\left(\vec{a}\right) = \hat{\vec{z}}_{j} + \vec{a}, \quad \hat{U}^{\dagger}\left(\mathcal{R}\right) \hat{\vec{z}}_{j} \hat{U}\left(\mathcal{R}\right) =  \mathcal{R} \hat{\vec{z}}_{j},
\label{eq:SymmetryEq_5}
\end{equation}
for all $j = 1, \ldots, N$ and $\mathcal{R}$ is the rotation matrix. We have required that all meaningful observables, for a system with translational and rotational symmetries, but by Eq. \eqref{eq:SymmetryEq_5} this is not the case here and thus $\hat{\vec{z}} \notin \mathcal{A}_{phys}$ are not physically meaningful in the present case. On the other hand, we have
\begin{equation}
\hat{U}^{\dagger}\left(\vec{a}\right) \hat{\vec{p}}_{j} \hat{U}\left(\vec{a}\right) = \hat{\vec{p}}_{j}, \quad \hat{U}^{\dagger}\left(\mathcal{R}\right) \hat{\vec{p}}_{j} \hat{U}\left(\mathcal{R}\right) = \mathcal{R} \hat{\vec{p}}_{j},
\label{eq:SymmetryEq_6}
\end{equation}
which reveals why the invariance condition under the mentioned symmetries alone is not a sufficient condition to classify physically meaningful observables. The momentum operators $\hat{\vec{p}}_{j}$ are conjugate to non-physical positions $\hat{\vec{z}}_{j}$ so it is not reasonable to claim these observables meaningful, even though it is invariant under translations. We therefore need an additional condition to classify observables like kinetic energies constructed from absolute momenta as non-meaningful. To summarize now the requirements for a physically meaningful quantum mechanical theory of interest: i) States of the system are normalizable eigenstates of a Hamiltonian, ii) All observables are invariant under given symmetries of the full symmetry group, iii) Invariance of the observables under Galilean boosts. The last criterion can be used to filter out absolute momenta and those scalar observables that are invariant but non-meaningful, as in the case of absolute momenta and Hamiltonian $\hat{H}$. Overall, these invariance conditions require that the physical content of the theory be independent of the choice of reference frame.

We have now discussed the necessary items for us to be able to set up a postulate to guide us in finding meaningful observables. In the following, a subgroup of $G$ is denoted by $G'$, and we denote the group of Galilean boosts by $B$. Based on the physically justified criteria above, we need additional postulates to distinguish physically meaningful observables and we therefore postulate:
\begin{itemize}

\item \textbf{Postulate 1.} The set of all physically meaningful observables is:
\begin{equation}
\mathcal{A}_{phys} = \left\{\hat{O} \in \mathcal{A} : \hat{U}^{\dagger}\left(g\right) \hat{O} \hat{U}\left(g\right) = \hat{O}, \quad \forall g \in G', \quad \text{and} \quad \hat{U}^{\dagger}\left(b\right) \hat{O} \hat{U}\left(b\right) = \hat{O}, \quad  \forall b \in B\right\}.
\label{eq:SymmetryEq_4_1}
\end{equation}

\item \textbf{Postulate 2.} There exists a map:
\begin{equation}
\hat{M}: \mathcal{A} \rightarrow \mathcal{A}_{phys}.
\label{eq:SymmetryEq_4_2}
\end{equation}

\end{itemize}
\noindent
Now an important question remains: what $G'$ actually is, given the Hamiltonian. In other words, what are actually the symmetries of the system that can be considered as describing the same physical situation. We claim here that global translations and rotations are included in $G'$, but there may be other relevant ones, like the permutation symmetry of identical particles \cite{Dirac-PrinciplesOfQM-1958}. In this work we concentrate on translations and rotations only, these symmetries originate from the underlying non-relativistic spacetime structure. The invariance requirement similar to Postulate 1 is given also elsewhere in the literature on quantum reference frames and relational quantum mechanics \cite{Loveridge-SymmetryReferenceFramesAndRelationalQuantitiesInQuantumMechanics-2018}, but the further restriction introduced here is given by the boost invariance, not in the symmetry group of the Hamiltonian. The requirement of boost invariance is natural, since non-invariant observables depend on the choice of inertial frame of reference, which contradicts our understanding of how nature works. Postulate 2 is physically justified: there must exist a map from the full set of observables to the set of physically meaningful observables, since otherwise a physically meaningful description of the system could not in general be constructed. These postulates allow a starting point for identification of physically meaningful observables. The process of obtaining the observables in $\mathcal{A}_{phys}$ of Postulate 1 from those that are not in $\mathcal{A}_{phys}$, is actually well known and it is called the reduction.

We now have all the necessary tools to proceed and discuss the reduction in more detail, namely finding the map $\hat{M}$ and thereby transforming the problem so that it is expressed in terms of the physically meaningful observables postulated above, while projecting out those that are not physically meaningful. In the following, we outline the reduction process and consider translational and rotational reduction separately.

\section{Reduction}
\label{Reduction}

We start by considering position and momentum observables. These are very central observables for our current problem as the Hamiltonian is a function of these observables, that is, if we can find physically meaningful position and momentum observables, we can write down a physically meaningful Hamiltonian as a function of them. To obtain the physically meaningful observables $\left(\hat{\vec{z}}',\hat{\vec{p}}'\right)$ from the original observables $\left(\hat{\vec{z}},\hat{\vec{p}}\right)$, we need to find a suitable map of Postulate 2, namely
\begin{equation}
\hat{M}: \left(\hat{\vec{z}},\hat{\vec{p}}\right) \rightarrow \left(\hat{\vec{z}}',\hat{\vec{p}}'\right).
\label{eq:ReductionEq_1}
\end{equation}
The question then is, what are the allowed forms of $\hat{M}$ and what properties are required to meet the Postulates 1 and 2 of Sec. \ref{Symmetry}. We write formally the map as the following composition
\begin{equation}
\hat{M} = \hat{\Pi}_{r} \circ \hat{C}_{r} \circ \hat{\Pi}_{t} \circ \hat{C}_{t},
\label{eq:ReductionEq_2}
\end{equation}
where $\hat{C}_{r},\hat{C}_{t}$ are unitary maps for rotation and translation, respectively, and $\hat{\Pi}_{r},\hat{\Pi}_{t}$ the corresponding projections. We note that while $\hat{C}_{r},\hat{C}_{t}$ are invertible, the projections $\hat{\Pi}_{r},\hat{\Pi}_{t}$ are not. Thus $\hat{M}$ cannot be invertible and we also note that it is not unique. We can write the whole transformation as
\begin{equation}
(\hat{\vec{z}},\hat{\vec{p}}) \xrightarrow{\,\hat{C}_{t}\,} (\hat{\vec{Z}}_{cm},\hat{\vec{P}}_{cm}, \hat{\vec{z}}_{t},\hat{\vec{p}}_{t}) \xrightarrow{\,\hat{\Pi}_{t}\,} (\hat{\vec{z}}_{t},\hat{\vec{p}}_{t}) \xrightarrow{\,\hat{C}_{r}\,} (\hat{\boldsymbol{\Omega}},\hat{\vec{p}}_{\Omega}, \hat{\vec{z}}_{s},\hat{\vec{p}}_{s}) \xrightarrow{\,\hat{\Pi}_{r}\,} (\hat{\vec{z}}',\hat{\vec{p}}'),
\label{eq:ReductionEq_3}
\end{equation}
where we denote the so-called orientation operators by $\hat{\boldsymbol{\Omega}}$ and the so-called shape operators by $\hat{\vec{z}}_{s}$. Moreover $(\hat{\vec{z}}_{t},\hat{\vec{p}}_{t})$ are assumed to be translational invariant observables and $(\hat{\vec{z}}_{s},\hat{\vec{p}}_{s})$ rotationally invariant. The purpose of the unitary parts is to transform the operators to those built from the invariant ones and the corresponding states. The unitary transformations transform the non-meaningful observables to meaningful ones, but there necessarily always remains non-invariant variables like $(\hat{\vec{Z}}_{cm},\hat{\vec{P}}_{cm})$. The projectors are needed for projecting out the unphysical degrees of freedom, the redundant degrees of freedom that are not group invariants. For example separating out the center-of-mass coordinates from the internal ones. The projectors on operators are thus maps of the form
\begin{align}
\hat{\Pi}_{t}& :  \hat{O}_{t}(\hat{\vec{Z}}_{cm},\hat{\vec{P}}_{cm}, \hat{\vec{z}}_{t},\hat{\vec{p}}_{t}) \rightarrow  \hat{O}_{phys}(\hat{\vec{z}}_{t},\hat{\vec{p}}_{t}), \nonumber \\
\hat{\Pi}_{r}& :  \hat{O}_{r}(\hat{\boldsymbol{\Omega}},\hat{\vec{p}}_{\Omega}, \hat{\vec{z}}_{s},\hat{\vec{p}}_{s}) \rightarrow  \hat{O}_{phys}(\hat{\vec{z}}',\hat{\vec{p}}').
\label{eq:ReductionEq_4}
\end{align}
In particular the physically meaningful Hamiltonian is $\hat{H}'(\hat{\vec{z}}',\hat{\vec{p}}')$ and now the question is, what is the general form of the Hamiltonian and the transformed and projected variables $(\hat{\vec{z}}',\hat{\vec{p}}')$. The unitary maps can be written as \cite{Wagner-UnitaryTransformationsInSolidStatePhysics-1986}
\begin{equation}
\hat{C}_{t} = e^{\hat{S}_{t}}, \quad \hat{C}_{r}  = e^{\hat{S}_{r}},
\label{eq:ReductionEq_5}
\end{equation}
and as $\hat{C}_{t}$ acts on $\left(\hat{\vec{z}},\hat{\vec{p}}\right)$, the quantity $\hat{S}_{t}$ must be a function of $\hat{\vec{z}}$ and $\hat{\vec{p}}$. In a similar way for $S_{r}$, and thus
\begin{equation}
\hat{S}_{t} = \hat{S}_{t}(\hat{\vec{z}},\hat{\vec{p}}), \quad \hat{S}_{r}  = \hat{S}_{r}(\hat{\vec{z}}_{t},\hat{\vec{p}}_{t}).
\label{eq:ReductionEq_6}
\end{equation}
Postulates 1 and 2, together with the map $\hat{M}$, define an approach seemingly analogous to geometric quotient-based reduction \cite{Rieffel-InducedRepresentationsOfCalgebras-1974,Landsman-MathematicalTopicsBetweenClassicalAndQuantumMechanics-1998}. However, we do not claim a general equivalence and it remains to be proved. The unitary parts of the map $\hat{M}$ can be interpreted as changes of quantum reference frames \cite{Aharonov-QuantumFramesOfReference-PhysRevD.30.368-1984,Giacomini-QuantumMechanicsAndTheCovarianceOfPhysicalLawsInQuantumReferenceFrames-2019}. That is, the transformation to a new set of invariant observables is a quantum reference frame transformation, but still not sufficient for us to obtain physically meaningful observables as we need projections to neglect the non-meaningful observables, which are necessarily present after a unitary transformation and this process cannot be achieved by unitary transformation only.

We then ask what are the allowed forms of physically meaningful observables. We start from the translational symmetry and if the system had only translational symmetry the physically meaningful quantities would be invariants of the translational group and boosts. In this case, the absolute position operators $\hat{\vec{z}}$, which are not invariant under translations, are built on affine Euclidean space, a space of points. The invariants thus belong to a different space that is the vector space of displacements associated with the Euclidean space. Concretely, the unitary transformation can be chosen such that it corresponds to a linear transformation
\begin{equation}
\hat{C}^{\dagger}_{t} \hat{\vec{z}}_{i} \hat{C}_{t} = \sum_{j} T_{ij} \hat{\vec{z}}_{j} = \hat{\vec{z}}_{ti}, \ i \neq n, \quad \hat{C}^{\dagger}_{t} \hat{\vec{z}}_{n} \hat{C}_{t} = \sum_{j} T_{nj} \hat{\vec{z}}_{j} = \hat{\vec{Z}}_{cm},
\label{eq:ReductionEq_7}
\end{equation}
such that apart from the center-of-mass coordinate $\hat{\vec{Z}}_{cm}$, the operators $\hat{\vec{z}}_{ti}$ are differences of operators built from $\hat{\vec{z}}_{j}$. For example, the quantity $\hat{\vec{z}}_{i} - \hat{\vec{z}}_{j}$ is invariant under translations. Now further conditions on $T_{ij}$, and thus on $\hat{S}_{t}$, can be set to ensure that one of the operators $\hat{\vec{z}}_{ti}$ is the center-of-mass position operator $\hat{\vec{Z}}_{cm}$ of the system and the remaining operators are invariants of the translational group, and functions of at least two different observables. Given the system also has rotational symmetry, the variables $(\hat{\vec{z}}_{t},\hat{\vec{p}}_{t})$ are still not physically meaningful and we must continue to further transform these observables to invariants of the rotational group.

The invariants of the rotational group are different kinds of mathematical objects. First, the physically meaningful operators in $(\hat{\vec{z}}',\hat{\vec{p}}')$ are functions of $(\hat{\vec{z}}_{t},\hat{\vec{p}}_{t})$, but the components appearing in $\hat{\vec{z}}'$ cannot be components of vectors. Namely, a vector cannot be invariant under rotations. For this reason, the unitary cannot correspond to linear transformations on $(\hat{\vec{z}}_{t},\hat{\vec{p}}_{t})$, but has to be of a form to produce a non-linear transformation instead. This means a more complicated functional dependence in $\hat{S}_{r}(\hat{\vec{z}}_{t},\hat{\vec{p}}_{t})$. From the requirement of rotational invariance it thus follows that the variables transform in a more complicated manner in $\hat{C}_{r}$. It turns out \cite{Littlejohn-GaugeFieldsInTheSeparOfRotatAndIntMotionsIntheNbodyProb-RevModPhys.69.213-1997} that we cannot project out three degrees of freedom from all observables, like the Hamiltonian, but there will be, in general, one invariant orientation variable which cannot be projected out exactly. There will be coupling between the rotational invariant internal variables and the invariant orientation variable. The invariants of the rotational group can be quantities like $\hat{\vec{z}}_{ti} \cdot \hat{\vec{z}}_{tj}$, $\hat{\vec{z}}_{ti} \cdot (\hat{\vec{z}}_{tj} \times \hat{\vec{z}}_{tj'})$ and functions alike.

After choosing the transformation $\hat{M}$ giving the observables respecting our postulates, the transformed Schrödinger equation is
\begin{equation}
\hat{H}' \ket{\Psi'} = E' \ket{\Psi'}.
\label{eq:ReductionEq_8}
\end{equation}
An interesting consequence of Postulates 1–2 is that the physically meaningful observables are not associated with individual original non-invariant operators taken in isolation. Rather, as discussed above, they are constructed from invariant combinations of the original non-invariant operators and are therefore relational in character. In this sense, physically meaningful observables are always defined relative to others. This is in line with the ideas of relational quantum mechanics \cite{Zurek-EnvironmentInducedSuperselectionRules-PhysRevD.26.1862-1982,Rovelli-RelationalQuantumMechanics-1996,Loveridge-SymmetryReferenceFramesAndRelationalQuantitiesInQuantumMechanics-2018}, and for this reason we refer to the reduced theories based on the Hamiltonian of Eq. \eqref{eq:ReductionEq_8} as relational theories \cite{Harkonen-BeyondBornOppenheimerGreensFunctionTheoriesAbsoluteAndRelational-2025}.

Up to this point, we have aimed at formulating the problem as physically meaningful. A special case of such a process was done in the very early stages of quantum mechanics: first in solving the Coulomb problem of hydrogen \cite{Schrodinger-QuantisierungAlsEigenwertproblemErsteMitteilung-1926,Schrodinger-QuantisierungAlsEigenwertproblemZweiteMitteilung-1926} and by Born and Oppenheimer in considering the Coulomb problem for molecules \cite{Born-OppenheimerAdiabaticApprox-1927}. Both cases likely followed to some extent the literature of classical mechanics and how many-body problems in those areas were actually solved by exploiting reduction. Therefore it is interesting to note that methods known already before quantum mechanics seem to play a rather central role in the quantum mechanical many-body case and can be connected \cite{Harkonen-BeyondBornOppenheimerGreensFunctionTheoriesAbsoluteAndRelational-2025} to many very modern and recent areas of physics research. Now that we take Eq. \eqref{eq:ReductionEq_8} physically meaningful, we may ask how to solve it in practice. The state size grows exponentially as a function of particle number rendering the equation impossible to solve even in the case of rather small molecules. Thus we essentially face the same problem of complexity whether or not the theory is relational and more efficient approaches to actually compute the observables are needed. One central simplification almost always assumed is the so-called Born-Oppenheimer (BO) approximation \cite{Born-OppenheimerAdiabaticApprox-1927,Born-Huang-DynamicalTheoryOfCrystalLattices-1954}, which underlies many of the approaches and hides some of the problems related to symmetries. It turns out that some of these approximate theories need not necessarily be reduced to relational ones due to broken symmetries. To some extent, this applies also to the case of conventional density functional theory (DFT) \cite{Hohenberg-DFT-PhysRev.136.B864-1964,KohnSham-DFT-PhysRev.140.A1133-1965,DreizlerGross-DFTbook-1990}. As a side note, the so-called multicomponent DFT, where the Hamiltonian $H$ is used as a starting point, is formulated as absolute \cite{Gidopoulos-KohnShamEquationsForMulticomponentSystemsTheExchangeAndCorrelationEnergyFunctional-PhysRevB.57.2146-1998} and also as relational \cite{Kreibich-MulticompDFTForElectronsAndNuclei-PhysRevLett.86.2984-2001,Kreibich-MulticompDFTForElectronsAndNuclei-PhysRevA.78.022501-2008}. In Sec. \ref{ManyBodyGreensFunctionTheories} we summarize some central aspects of the BO approximation and how it relates to our discussion so far. We also connect our discussions to the field theoretic many-body Green's function approach \cite{Schwinger-OnTheGreensFunctionsOfQuantizedFieldsI-1951,Gross-ManyParticleTheory-1991}.

\section{Many-Body Green's Function Theories}
\label{ManyBodyGreensFunctionTheories}

The system we consider here comprises $N_{e}$ electrons and $N_{n}$ nuclei, the total particle number is thus $N = N_{e} + N_{n}$. We denote all electron position operators in the position representation as $\vec{r} \equiv \vec{r}_{1}, \ldots, \vec{r}_{N_{e}}$ and in a similar way the nuclear operators as $\vec{R} \equiv \vec{R}_{1}, \ldots, \vec{R}_{N_{n}}$. The Hamiltonian operator is a special case of Eq. \eqref{eq:SymmetryEq_1} and can be written as
\begin{equation}
H = T_{e} + T_{n} + V_{ee} + V_{en} + V_{nn},
\label{eq:ManyBodyGreensFunctionTheoriesEq_1}
\end{equation}
where $T_{e}$ is the kinetic energy for electrons, $T_{n}$ is the kinetic energy for nuclei, $V_{ee}$, $V_{en}$ and $V_{nn}$, the Coulomb potential energies for electron-electron, electron-nucleus and nucleus-nucleus interactions, respectively. This Hamiltonian satisfies Eq. \eqref{eq:SymmetryEq_2}, $U^{\dagger}\left(g\right) H U\left(g\right) = H, \quad g \in G$. The time-independent Schrödinger equation, is $H \Psi\left(\vec{r},\vec{R}\right) = E \Psi\left(\vec{r},\vec{R}\right)$, but we have already discussed that the eigenstates satisfying this equation are not normalizable and thus not physically relevant in describing molecules and solids. At the same time, the whole theory of solids \cite{Born-Huang-DynamicalTheoryOfCrystalLattices-1954} and the many-body Green's function theories \cite{Baym-field-1961,Giustino-ElectronPhononInteractFromFirstPrinc-RevModPhys.89.015003-2017,Stefanucci-InAndOutOfEquilibriumAbInitioTheoryOfElectronsAndPhonons-PhysRevX.13.031026-2023} are based on Eq. \eqref{eq:ManyBodyGreensFunctionTheoriesEq_1} as a starting point. To see why such theories can still be formulated to provide information about the systems discussed here can be justified as follows. First simplification is the Born-Oppenheimer approximation \cite{Born-Huang-DynamicalTheoryOfCrystalLattices-1954} such that $\Psi\left(\vec{r},\vec{R}\right) \approx \Phi_{\vec{R}}\left(\vec{r}\right) \chi\left(\vec{R}\right)$, where the electronic $\Phi_{\vec{R}}\left(\vec{r}\right)$ and nuclear wave functions $\chi\left(\vec{R}\right)$ satisfy
\begin{align}
H_{BO} \Phi_{\vec{R}}\left(\vec{r}\right) &= \epsilon_{BO}\left(\vec{R}\right) \Phi_{\vec{R}}\left(\vec{r}\right), \label{eq:ManyBodyGreensFunctionTheoriesEq_2_1} \\
H_{n} \chi\left(\vec{R}\right) &= E \chi\left(\vec{R}\right), \label{eq:ManyBodyGreensFunctionTheoriesEq_2_2}
\end{align}
where $H_{BO} = H - T_{n}$ and $H_{n} = T_{n} + \epsilon_{BO}\left(\vec{R}\right)$. When the electronic problem is treated independently, the nuclei variables can be treated as parameters and this breaks the spatial symmetries. Namely, the Hamiltonian is not in general invariant under 
\begin{equation}
U^{\dagger}_{e}\left(g\right) H_{BO} U_{e}\left(g\right) \neq H_{BO}, \quad g \in G' \leq G,
\label{eq:ManyBodyGreensFunctionTheoriesEq_3}
\end{equation}
and thus $H_{BO}$ when considered in electronic space only. Here we denote $U_{e}\left(g\right)$ symmetry unitary acting on electronic variables only, which are the only quantum mechanical variables when Eq. \eqref{eq:ManyBodyGreensFunctionTheoriesEq_2_1} is treated as an independent equation. Instead the Hamiltonian can have discrete point group and space group symmetries with respect to parametric nuclear variables. Thus, the electronic BO approximation alone breaks the spatial symmetries and allows reasonable eigenstates without reduction. This is also the reason why DFT and the field theoretic electronic many-body Green's function approach assuming BO approximation do not suffer from the mentioned symmetry related issues. When we go back to the full problem, even the approximate one, $\Psi\left(\vec{r},\vec{R}\right) \approx \Phi_{\vec{R}}\left(\vec{r}\right) \chi\left(\vec{R}\right)$, then the symmetry related aspects return, as has been discussed \cite{Harkonen-ExactFactorizationOfTheManyBodyGreensFunctionTheoryOfElectronsAndNuclei-PhysRevB.106.205137-2022}. Moreover, the nuclei problem of Eq. \eqref{eq:ManyBodyGreensFunctionTheoriesEq_2_2} still has the translational and rotational symmetries \cite{Born-Huang-DynamicalTheoryOfCrystalLattices-1954}. Namely, the translational symmetry leads to the so-called acoustic sum rule ensuring that there will be three vibrational modes of zero frequency \cite{Born-Huang-DynamicalTheoryOfCrystalLattices-1954} and we can actually explicitly see in the harmonic approximation for the nuclei that this indeed renders the eigenstates $\chi\left(\vec{R}\right)$ physically non-meaningful as such \cite{Harkonen-ExactFactorizationOfTheManyBodyGreensFunctionTheoryOfElectronsAndNuclei-PhysRevB.106.205137-2022,Harkonen-BeyondBornOppenheimerGreensFunctionTheoriesAbsoluteAndRelational-2025}. The rotational symmetry is also present, but it is broken in practical approaches by imposing the Born-von Karman periodic boundary conditions \cite{Born-Huang-DynamicalTheoryOfCrystalLattices-1954}. In some cases, like position operator expected values in crystals, the remaining necessary center-of-mass separation can be dealt with simply by neglecting the zero wave vector $\vec{q} = 0$ acoustic modes of nuclear vibrations \cite{Harkonen-NTE-2014,Harkonen-BreakdownOfTheBornOppenheimerApproximationInSolidHydrogenAndHydrogenRichSolids-Arxiv-2023,Harkonen-OnTheBreakdownOfTheBornOppenheimerApproximationInLiHandLiD-2026}.

The many-body Green's function approaches are not disconnected from these considerations as pointed out \cite{Harkonen-ManyBodyGreensFunctionTheoryOfElectronsAndNucleiBeyondTheBornOppenheimerApproximation-PhysRevB.101.235153-2020,Harkonen-ExactFactorizationOfTheManyBodyGreensFunctionTheoryOfElectronsAndNuclei-PhysRevB.106.205137-2022,Harkonen-BeyondBornOppenheimerGreensFunctionTheoriesAbsoluteAndRelational-2025}. Namely, the beyond-BO Green's functions (zero temperature here for simplicity) can be defined
\begin{equation} 
G\left(\vec{y}t,\vec{y}'t'\right) \equiv -i \braket{\Psi|\mathcal{T} \{ \hat{\psi}\left(\vec{y}t\right) \hat{\psi}^{\dagger}\left(\vec{y}'t'\right) \}|\Psi},
\label{eq:ManyBodyGreensFunctionTheoriesEq_4}
\end{equation}
where we denote the time-ordering by $\mathcal{T}\{ \cdots \}$ and $\ket{\Psi}$ is some eigenstate of the Hamiltonian as we are here only interested in stationary states. The time evolution in the electronic field operators $\hat{\psi}\left(\vec{y}t\right)$ and $\hat{\psi}^{\dagger}\left(\vec{y}'t'\right)$ is defined with respect to the Hamiltonian $\hat{H}$. However, we have already justified that these states are not normalizable and thus the Green's function of Eq. \eqref{eq:ManyBodyGreensFunctionTheoriesEq_4} does not give physically meaningful observables. The BO electronic Green's function on the other hand \cite{Harkonen-ExactFactorizationOfTheManyBodyGreensFunctionTheoryOfElectronsAndNuclei-PhysRevB.106.205137-2022}
\begin{equation} 
G^{BO}_{\vec{R}}\left(\vec{y}t,\vec{y}'t'\right) \equiv -i \braket{\Phi_{\vec{R}}|\mathcal{T} \{ \hat{\psi}\left(\vec{y}t\right) \hat{\psi}^{\dagger}\left(\vec{y}'t'\right) \}|\Phi_{\vec{R}}}.
\label{eq:ManyBodyGreensFunctionTheoriesEq_5}
\end{equation}
provides meaningful observables as the symmetries are broken, as discussed above. In Eq. \eqref{eq:ManyBodyGreensFunctionTheoriesEq_5}, the time evolution in the electronic field operators is now defined with respect to the Hamiltonian $\hat{H}_{BO}$. However, if we need a beyond-BO many-body Green's function theory we need to formulate the theory as relational, namely based on the Hamiltonian $\hat{H}'$ of Eq. \eqref{eq:ReductionEq_8}. That is
\begin{equation} 
G'\left(\vec{y}t,\vec{y}'t'\right) \equiv -i \braket{\Psi'|\mathcal{T} \{ \hat{\psi}\left(\vec{y}t\right) \hat{\psi}^{\dagger}\left(\vec{y}'t'\right) \}|\Psi'},
\label{eq:ManyBodyGreensFunctionTheoriesEq_6}
\end{equation}
where the time evolution in the electronic field operators is defined with respect to the Hamiltonian $\hat{H}'$ and $\ket{\Psi'}$ is some eigenstate of this Hamiltonian as we are here only interested in stationary states. This is exactly what is done in the reduced/relational many-body Green's function approaches \cite{vanLeeuwen-FirstPrincElectronPhonon-PhysRevB.69.115110-2004,Harkonen-ManyBodyGreensFunctionTheoryOfElectronsAndNucleiBeyondTheBornOppenheimerApproximation-PhysRevB.101.235153-2020,Harkonen-ExactFactorizationOfTheManyBodyGreensFunctionTheoryOfElectronsAndNuclei-PhysRevB.106.205137-2022,Harkonen-BeyondBornOppenheimerGreensFunctionTheoriesAbsoluteAndRelational-2025} and in these approaches the Green's functions defined with respect to $\ket{\Psi'}$ and $\hat{H}'$ following from a particular choice of map $\hat{M}$ of Postulate 2.

\section{Conclusions}
\label{Conclusions}

In this work, we have considered symmetry and related reduction in many-body quantum mechanical systems from different points of view. We gave justifications why reduction is necessary and leads to relational theories given the system has certain symmetries. Moreover, we formulate a unified postulate-based approach for obtaining a physically meaningful quantum many-body theory of normalizable stationary states from a given Hamiltonian. This postulate-based approach provides a unified framework connecting \cite{Harkonen-BeyondBornOppenheimerGreensFunctionTheoriesAbsoluteAndRelational-2025} several aspects of classical and quantum theory. The postulates ensure that the theory is formulated in terms of physically meaningful observables and several consequences follow. First, symmetries, and the associated superselection rules \cite{Aharonov-ChargeSuperselectionRule-PhysRev.155.1428-1967,Page-EvolutionWithoutEvolutionDynamicsDescribedByStationaryObservables-PhysRevD.27.2885-1983,Aharonov-QuantumParadoxesQuantumTheoryForThePerplexed-2005}, allow us to identify non-meaningful observables. Second, reduction theory \cite{Rieffel-InducedRepresentationsOfCalgebras-1974,Landsman-MathematicalTopicsBetweenClassicalAndQuantumMechanics-1998,Littlejohn-GaugeFieldsInTheSeparOfRotatAndIntMotionsIntheNbodyProb-RevModPhys.69.213-1997,Marsden-MechanicalSystemsSymmetryAndReduction-2009} provides a systematic way to transform these non-meaningful observables into meaningful ones. Third, as suggested in this work, the unitary part of this transformation can be interpreted as a change of quantum reference frame \cite{Aharonov-QuantumFramesOfReference-PhysRevD.30.368-1984,Giacomini-QuantumMechanicsAndTheCovarianceOfPhysicalLawsInQuantumReferenceFrames-2019}. Fourth, reduction leads to a separation of different types of motion, such as translational, orientational, and internal degrees of freedom \cite{Born-OppenheimerAdiabaticApprox-1927,Watson-SimplificationOfTheMolecularVibrationRotationHamiltonian-1968,Sutcliffe-TheDecouplingOfElectronicAndNuclearMotions-2000}. Finally, the physically meaningful observables that remain after reduction are necessarily relational, consistent with the relational interpretation of quantum mechanics \cite{Zurek-EnvironmentInducedSuperselectionRules-PhysRevD.26.1862-1982,Rovelli-RelationalQuantumMechanics-1996,Poulin-ToyModelForARelationalFormulationOfQuantumTheory-2006,Loveridge-SymmetryReferenceFramesAndRelationalQuantitiesInQuantumMechanics-2018,Calosi-RelationalQuantumMechanicsAtTheCrossroads-2024}.

Compared with existing relational approaches, our formulation additionally requires invariance of physical observables under Galilean boosts. This requirement is natural from a physical perspective, since it expresses the absence of a preferred inertial frame. In this sense, it is the Galilean analogue of the relativity principle, according to which the laws of physics take the same form in all inertial frames \cite{Einstein-ZurElektrodynamikBewegterKorper-1905}. Absolute quantities do not satisfy this requirement, since they are not invariant under boosts. We therefore argue that boost invariance should be imposed as an additional criterion, beyond invariance under the chosen subgroup of symmetries, when identifying the physically meaningful observables.

This work makes efforts to identify the general structure underlying symmetry-related reductions, their consequences, and the minimal set of constraints required to formulate physically meaningful quantum many-body theories. At the same time, we interpret many theoretical constructs as important elements in constructing a quantum theory of physically meaningful relational observables. We hope that the framework developed here helps to better connect different disciplines addressing related problems and provides a simple set of well-motivated necessary principles for achieving a deeper understanding of nature.

\printbibliography

\end{document}